\documentclass[aps,prb,reprint,superscriptaddress,longbibliography,floatfix]{revtex4-2}
\usepackage{graphicx,amsmath,amssymb,bm,xcolor,url,verbatim}
\usepackage[colorlinks=true,citecolor=blue,linkcolor=blue,urlcolor=blue]{hyperref}
\begin{document}

\title{Switching-Current Signatures of Magnetically Active Avalanche-like Metastability in InAs/Al Nanowire Josephson Junctions}

\author{O. Durante}
\email{odurante@unisa.it}
\thanks{Corresponding author}
\affiliation{Dipartimento di Fisica ``E. R. Caianiello'', Universit\`a di Salerno, Fisciano (SA), Italy}

\author{R. Citro}
\email{rocitro@unisa.it}
\thanks{Corresponding author}
\affiliation{Dipartimento di Fisica ``E. R. Caianiello'', Universit\`a di Salerno, Fisciano (SA), Italy}
\affiliation{CNR-SPIN, c/o Universit\`a di Salerno, I-84084 Fisciano (Salerno), Italy}
\affiliation{INFN - Gruppo Collegato di Salerno, Fisciano (SA), Italy}

\author{E. Strambini}
\affiliation{Istituto Nanoscienze – CNR, NEST-SNS, Pisa, Italy}

\author{V. Demontis}
\affiliation{Dipartimento di Fisica, Universit\`a degli Studi di Cagliari, Monserrato, Cagliari, Italy}

\author{M. Rocci}
\affiliation{Istituto Nanoscienze – CNR, NEST-SNS, Pisa, Italy}

\author{A. Braggio}
\affiliation{Istituto Nanoscienze – CNR, NEST-SNS, Pisa, Italy}

\author{S. Battiato}
\affiliation{Dipartimento di Fisica e Astronomia ``E. Majorana'', Universit\`a di Catania, Catania, Italy}
\affiliation{CNR-IMM, Catania, Italy}

\author{V. Zannier}
\affiliation{Istituto Nanoscienze – CNR, NEST-SNS, Pisa, Italy}

\author{L. Sorba}
\affiliation{Istituto Nanoscienze – CNR, NEST-SNS, Pisa, Italy}

\author{F. Giazotto}
\affiliation{Istituto Nanoscienze – CNR, NEST-SNS, Pisa, Italy}

\author{C. Guarcello}
\email{cguarcello@unisa.it}
\thanks{Corresponding author}
\affiliation{Dipartimento di Fisica ``E. R. Caianiello'', Universit\`a di Salerno, Fisciano (SA), Italy}
\affiliation{INFN - Gruppo Collegato di Salerno, Fisciano (SA), Italy}

%%%%%%%%%%%%%%%%%%%%%%%%%%%%%%%%%%%%%%%%%%%%%%%%%%%%%%%%%%%%%%%%%%%%%
%% Abstract
%%%%%%%%%%%%%%%%%%%%%%%%%%%%%%%%%%%%%%%%%%%%%%%%%%%%%%%%%%%%%%%%%%%%%
\begin{abstract}
We report switching-current signatures of avalanche-like metastable reconfigurations in an $n$-doped InAs/Al nanowire Josephson junction. Under a perpendicular magnetic field, the device exhibits a low-field Fraunhofer-like modulation of the switching current together with discrete jumps emerging at $|B|\approx 3$~mT. The jump field is nearly temperature independent from 30 to 900 mK, unlike the superconducting critical field. Combined with the reproducible sweep-direction dependence, this behavior points to avalanche-like reconfigurations of a metastable degree of freedom coupled to the weak link, with magnetic switching providing a possible microscopic interpretation. These results identify hybrid nanowire Josephson junctions as sensitive probes of magnetically active avalanche-like metastability at low magnetic fields.
\end{abstract}

\maketitle
%Published in Physical Review B. DOI: \href{https://doi.org/10.1103/sm95-vrgy}{10.1103/sm95-vrgy}.

\begin{figure*}[t]
 \centering
 \includegraphics[width=2\columnwidth]{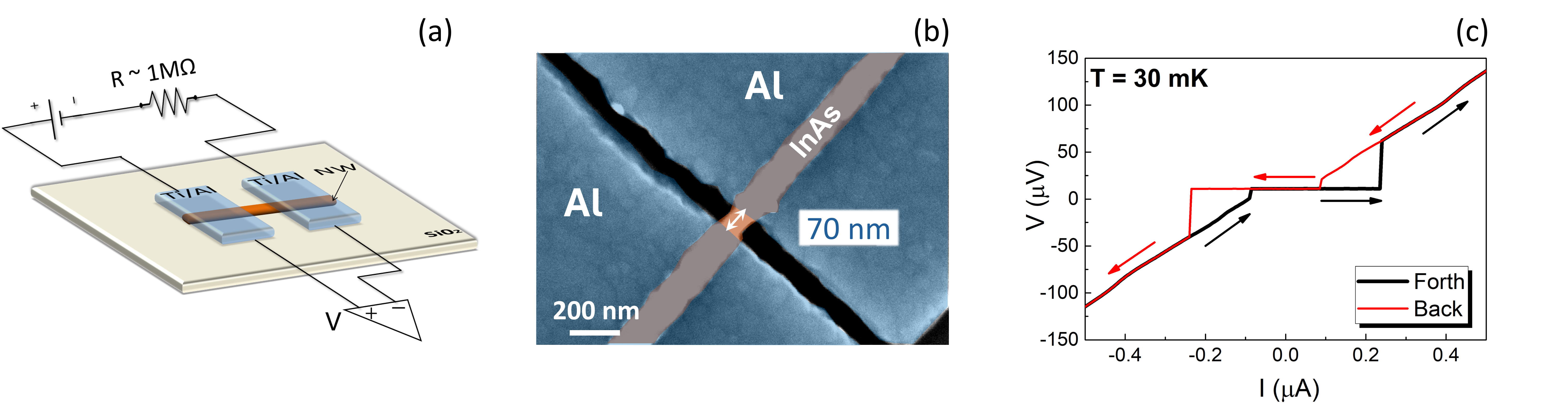}
\caption{(a) Schematic of the InAs/Al NW JJ and four-wire current-bias measurement setup. 
(b) False-colour SEM of the junction area showing the InAs NW contacted by two superconducting Al leads; interelectrode spacing $L\simeq70$~nm. 
(c) $V$--$I$ characteristic at $T=30$~mK showing a switching current $I_{\mathrm{sw}}\approx 0.25~\mu$A and hysteresis upon retrapping.}
 \label{fig:device}
\end{figure*}

Magnetic degrees of freedom in hybrid superconductor--semiconductor Josephson junctions (JJs) can profoundly affect superconducting transport, inducing phase shifts, nontrivial current--phase relations, and bistable responses.~\cite{Strambini2020,Szombati2016,Levajac2024,Razmadze2023} While such effects can be engineered in a controlled way by incorporating ferromagnetic elements or magnetically active barriers,~\cite{Kammermeier2024,Birge2024,Gharavi2017} intrinsic magnetic contributions, for instance from surface moments or correlated impurities in the hybrid environment, are more elusive, yet can strongly influence the Josephson response.~\cite{Strambini2020,Yang2021,Razmadze2023} Directly detecting and characterizing such magnetically active metastable configurations at the nanoscale remains challenging, particularly under cryogenic operating conditions, and their signatures can be intertwined with more conventional superconducting mechanisms.

Here, we identify a regime in which an InAs/Al nanowire (NW) JJ acts as a sensitive transport probe of avalanche-like reconfigurations of a magnetically active metastable degree of freedom, driven by a low perpendicular field and transduced into discrete switching-current jumps. We observe a Fraunhofer-like modulation of the switching current coexisting with reproducible jump events and pronounced sweep-direction dependence at fields of a few millitesla.
By tracking the field scale associated with the main switching-current discontinuity as a function of temperature, we find that it is essentially temperature independent. This behavior is difficult to reconcile with standard superconductivity-suppression mechanisms. It is more naturally described within a Barkhausen-like crackling-noise phenomenology, namely avalanche-like switching of a disordered metastable subsystem coupled to the junction, with magnetic reconfiguration as a natural microscopic interpretation.~\cite{Bozorth1932,Durin2006,Sethna2001}
 Within an effective-field picture, each reconfiguration changes a local field offset and may also reshape the Josephson response, thereby converting metastable magnetic switching into abrupt transport discontinuities.

Figure~\ref{fig:device} describes the InAs NW JJ and the corresponding four-wire current-bias measurement setup. Panel (a) shows the device layout and the current-bias configuration. The NWs, grown by Au-catalyzed chemical beam epitaxy with typical length $\sim 1.5~\mu$m and diameter $\sim 80$--$90$~nm,~\cite{Gomes2015,Demontis2019,Demontis2024} were $n$-doped using Se during growth. The NWs were deposited onto a degenerately doped Si substrate covered with a 300~nm SiO$_2$ insulating layer and contacted by Ti/Al (10/100~nm) electrodes defined by electron-beam lithography and evaporation. Prior to contact deposition, the NW surfaces were selectively cleaned with a diluted ammonium polysulfide, $(\mathrm{NH}_4)_2\mathrm{S}_x$, solution to remove the native oxide and passivate the interface. The scanning electron micrograph (SEM) of a representative junction is shown in panel (b), where the interelectrode spacing is $L \simeq 70$~nm. Across devices, the spacing defining the weak link typically ranges between 70 and 80~nm, although the effective junction length may differ slightly due to edge roughness and irregular Al coverage. Panel (c) shows a representative $V$--$I$ characteristic at $T=30$~mK, with a critical current of about 250~nA and a lower retrapping current, consistent with an underdamped junction.

A magnetic field $B$ was applied perpendicular to the device plane, modulating the Josephson response and providing a control parameter for possible magnetically active metastable degrees of freedom. Indeed, the weak link responds to a local effective field that can differ substantially from the applied field due to flux focusing and stray-field contributions from nearby magnetization. Accordingly, slow field sweeps can reveal discrete reconfigurations of a magnetic subsystem coupled to the junction, giving rise to sweep-direction-dependent signatures, as further discussed in the Supplemental Material~\cite{SM}.

Figure~\ref{fig:interference}(a) shows the differential resistance map $dV/dI(I,B)$ at $T=30$~mK. The dark region corresponds to the zero-voltage superconducting state, while the bright boundary marks the transition to the dissipative branch with finite voltage. Throughout this work, we define the switching current $I_{\mathrm{sw}}(B)$ as the bias current at which the junction leaves the zero-voltage state (onset of dissipation) for a given $B$ and $T$. A key feature is an abrupt change in the behavior of $I_{\mathrm{sw}}(B)$ at a threshold field $|B|\approx 3$~mT. In particular, for $B_j^-<B<B_j^+$, the switching current seems to follow a Fraunhofer-like envelope, with $B_j^\pm$ defined as the fields at which discontinuities in $I_{\mathrm{sw}}(B)$ occur during quasi-adiabatic sweeps of $B$ from negative to positive values. Beyond these threshold values, i.e., for $B<B_j^-$ and $B>B_j^+$, the switching current profile exhibits a dense sequence of switching-current jumps, progressing toward supercurrent suppression. Notably, the right and left sides of the pattern differ, indicating a pronounced dependence of the junction response on the field-sweep history. These features reverse upon reversing the sweep direction.

At $T=30$~mK the device reaches $I_{\mathrm{sw}}(0)\approx0.24~\mu$A at zero field. At the first jump event, the change is $\Delta I_{\mathrm{sw}}(B_j^-)\approx0.13~\mu$A. The effect is robust over repeated sweeps, confirming that it is not a spurious fluctuation or noise artifact; moreover, we observed similar switching behavior in other devices fabricated under nominally identical conditions (see Sec.~S1 of Supplemental Material~\cite{SM}). Moreover, we tested symmetry under current inversion, ruling out a nonreciprocal Josephson-diode interpretation.~\cite{Nadeem2023,Ma2025}

Figure~\ref{fig:interference}(b) shows $I_{\mathrm{sw}}(B)$ for temperatures between 30 and 900~mK. The overall lobe-shaped modulation of $I_{\mathrm{sw}}(B)$ is compatible with a Josephson interference pattern. Indeed, near zero field, the envelope follows a Fraunhofer-like dependence, once flux focusing is properly taken into account (see Sec.~S2 of Supplemental Material~\cite{SM}). The threshold jump field remains close to $\pm3$~mT across the entire temperature range, whereas the overall supercurrent scale decreases with $T$ and vanishes around 1~K. To quantify these trends, we extract the temperature dependence of both the superconducting critical fields $B_c^\pm$, defined as the fields where $I_{\mathrm{sw}}\to 0$, and the jump fields $B_j^\pm$.

\begin{figure}[tbp]
 \centering
 \includegraphics[width=\columnwidth]{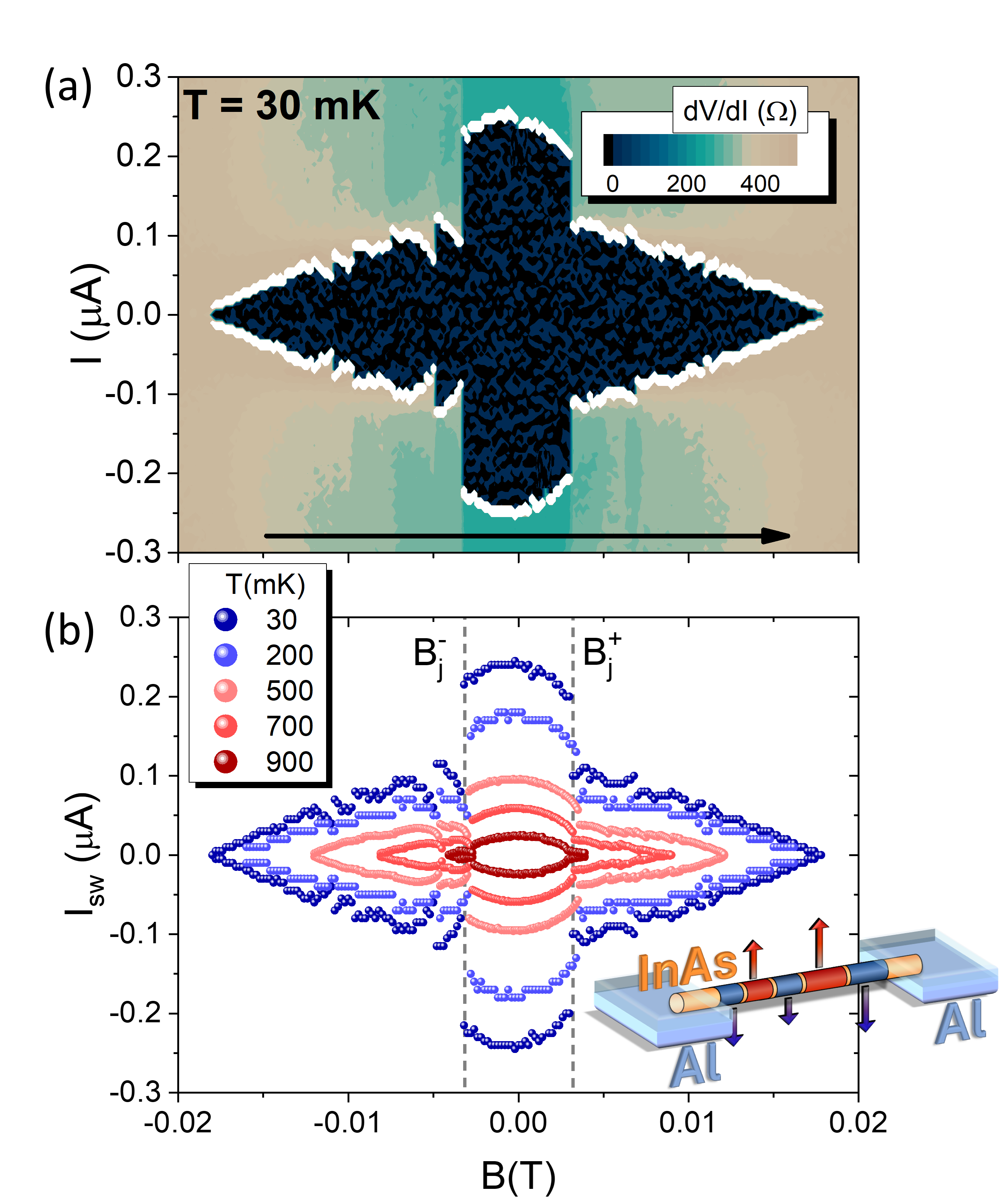}
\caption{(a) Differential resistance map $dV/dI(I,B)$ at $T=30$~mK. The dark region corresponds to the superconducting branch; the bright boundary marks $I_{\mathrm{sw}}(B)$. 
(b) Switching current $I_{\mathrm{sw}}(B)$ at different temperatures (30--900~mK), showing a Fraunhofer-like modulation and an abrupt transition at $|B|\approx3$~mT between two branches. The inset shows a cartoon of the InAs NW, between two superconducting Al leads, schematically represented as a sequence of magnetically active metastable regions, with two possible local configurations indicated in blue/red.}
 \label{fig:interference}
\end{figure}

\begin{figure}[tbp]
 \centering
 \includegraphics[width=0.8\columnwidth]{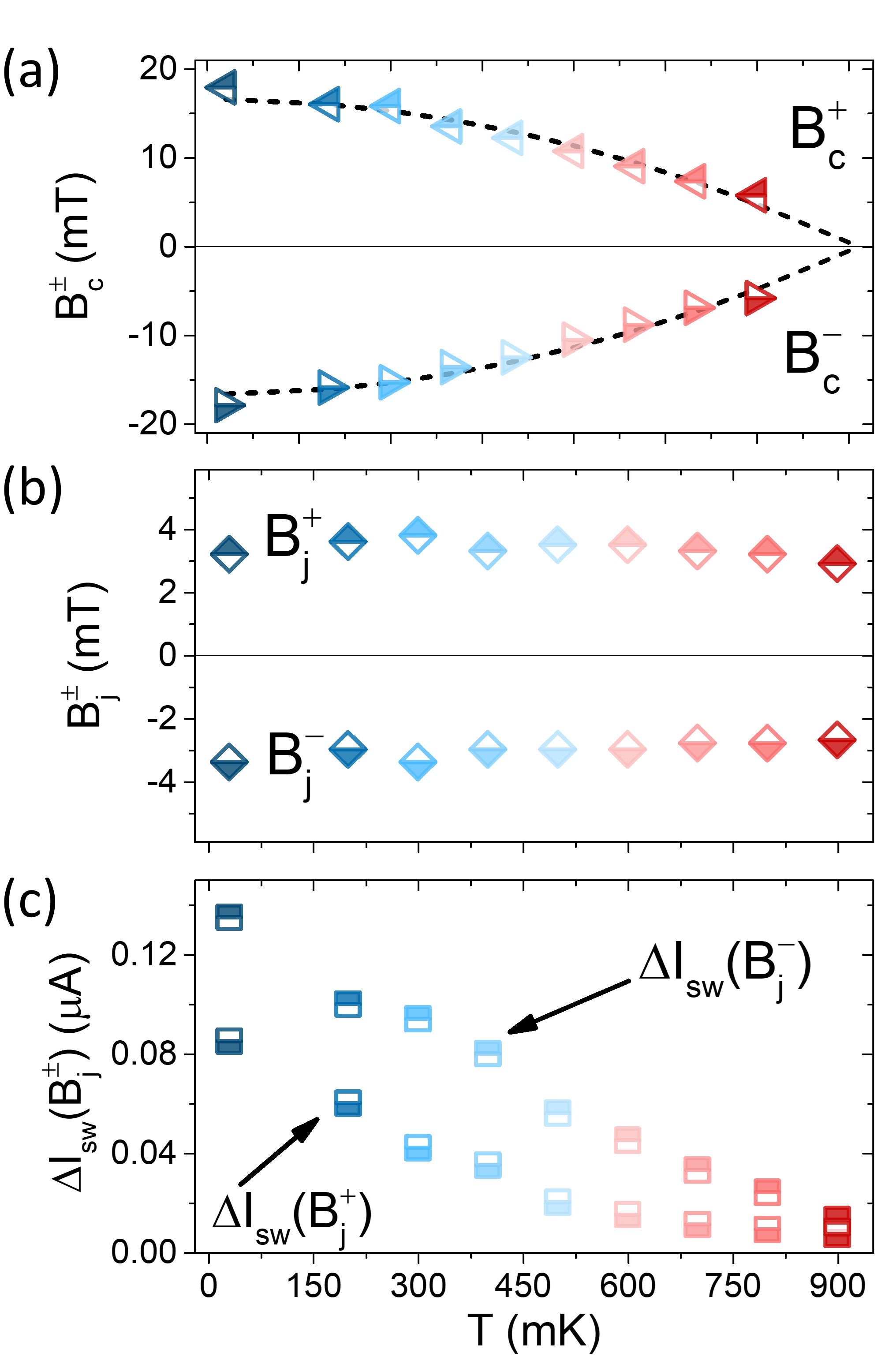}
\caption{(a) Positive and negative critical fields $B_c^\pm$ versus temperature; dotted lines are parabolic fits.
(b) Jump fields $B_j^\pm$ versus temperature, remaining nearly constant around $\pm3$~mT.
(c) Jump amplitude $\Delta I_{\mathrm{sw}}(B_j^\pm)$ versus temperature.}
 \label{fig:Tdep}
\end{figure}

As shown in Fig.~\ref{fig:Tdep}(a), the critical field $B_c^\pm(T)$ follows the expected thin-film behavior. A fit to $B_c(T)=B_c(0)\left[1-(T/T_c)^2\right]$ yields $B_c(0)=(16.6\pm0.4)$~mT and $T_c=(1.06\pm0.03)$~K for the positive branch. In contrast, $B_j^\pm$ is nearly temperature independent [Fig.~\ref{fig:Tdep}(b)]. This behavior is difficult to reconcile with simple vortex penetration or flux trapping in the superconducting leads, which would be expected to track temperature-dependent superconducting length and field scales. Furthermore, flux trapping/detrapping typically produces greater scan-to-scan variability in the positions of the switching features. 
Instead, the weak $T$ dependence of $B_j^\pm$ supports a threshold set by a magnetically active subsystem coupled to the weak link rather than by superconductivity suppression, a scenario further developed below.
The jump amplitude $\Delta I_{\mathrm{sw}}(B_j^\pm)$ decreases with $T$ but remains finite up to 900~mK [Fig.~\ref{fig:Tdep}(c)], consistent with robust bistability of the underlying magnetically active metastable configuration.

\begin{figure}[tbp]
 \centering
 \includegraphics[width=0.8\columnwidth]{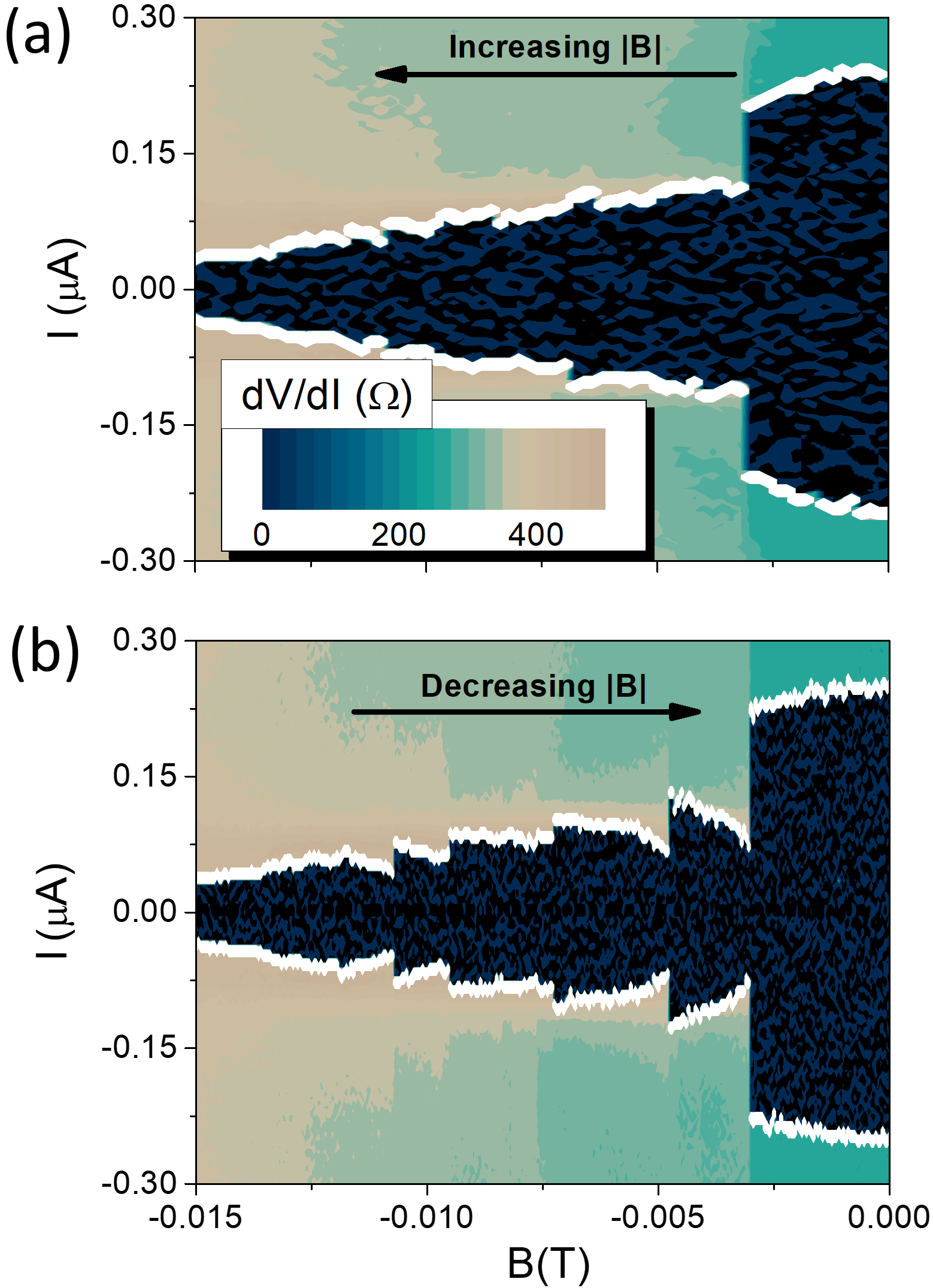}
\caption{Differential resistance maps $dV/dI(I,B)$ at $T=30$~mK for (a) a sweep from $B=0$ to $-15$~mT and (b) a sweep from $-15$~mT to $0$. 
Beyond the main switch near $|B|\simeq3$~mT, smaller discontinuities depend on sweep history, consistent with avalanche-like metastable switching of a magnetically active subsystem.} \label{fig:hist}
\end{figure}

The sweep-direction dependence is highlighted in Fig.~\ref{fig:hist}, which compares $dV/dI(I,B)$ maps acquired for opposite field sweeps at 30~mK. Beyond the main switch at $|B|\simeq3$~mT, additional smaller discontinuities appear in a reproducible, history-dependent manner. 
Such behavior is characteristic of metastability and is broadly compatible with a Barkhausen-like crackling-noise phenomenology, namely avalanche-like switching of a disordered metastable subsystem,~\cite{Bozorth1932,Durin2006,Sethna2001} where collective rearrangements of magnetically active regions or moments can lead to discrete changes in the local effective field acting on the junction.
This scenario is illustrated schematically in the inset of Fig.~\ref{fig:interference}(b), where a possible domain-like metastable structure in or near the NW contributes a local-field offset to the Josephson weak link.

Qualitatively similar switching-current steps were reported in engineered S-(S/F)-S superconducting switches incorporating a ferromagnetic finger, where stochastic domain depinning reshapes the stray-field landscape and locally quenches superconductivity in the constriction.~\cite{Kammermeier2024} 
Moreover, recent EuS-based hybrid structures provide a useful comparison in systems where the magnetic material is explicit. In full-shell InAs/EuS/Al NWs, transport combined with scanning SQUID magnetometry showed that superconductivity in the Al shell is correlated with reconfigurable EuS magnetic textures, appearing in multidomain configurations and disappearing in the saturated single-domain state.~\cite{Nandi2026} In Al/EuS bilayers, superconducting transport was shown to depend sensitively on the ratio between the superconducting coherence length and the magnetic-domain size, as first highlighted by tunneling spectroscopy measurements of the magnetic proximity effect,~\cite{Strambini2017} and later connected to percolative supercurrent flow when magnetic inhomogeneity is averaged over the superconducting length scale.~\cite{Maiani2025}
In our case, even in the absence of an engineered ferromagnetic layer, we observe reproducible low-field switching within a Fraunhofer-like pattern, consistent with discrete changes in a local intrinsic effective field.
This low-field metastable response should not be regarded as an established generic property of all InAs/Al NW JJs; its microscopic origin may depend on growth conditions, doping, surface treatment, interface quality, or device geometry.
This interpretation is consistent with independent measurements on nominally identical InAs/Al NWs,~\cite{Strambini2020,Yang2021} where a finite anomalous Josephson phase was attributed to the interplay of spin--orbit coupling and intrinsic magnetism, even in the absence of an external magnetic field.
In particular, Ref.~\cite{Strambini2020} reported a Kondo-like upturn of the normal-state resistance, consistent with dilute localized magnetic moments in the NW environment.
The anomalous phase-shift phenomenology discussed in Ref.~\cite{Strambini2020} is difficult to reconcile with trivial vortex entry,~\cite{Guarcello2020} but points to an intrinsic magnetic contribution that, combined with spin--orbit coupling in the NW, gives rise to an anomalous Josephson phase.~\cite{Strambini2020,Bergeret2015,Szombati2016,Assouline2019,Mayer2020,Shukrinov2022} 
Nontrivial structures in $I_{\mathrm{sw}}(B)$ have also been reported in other platforms, including InAs and Ge--Si NW junctions,~\cite{Tiira2017,Wu2024} and have been discussed in terms of alternative microscopic mechanisms such as parity transitions or orbital-selective transport.~\cite{Maiellaro2023,Guarcello2024,Maiellaro2024} In our device, the coexistence of a Fraunhofer-like envelope, pronounced sweep-direction dependence, and nearly temperature-independent jump positions is instead consistent with avalanche-like switching of a magnetically active metastable subsystem that abruptly modifies the local effective field.

A phenomenological description is obtained by distinguishing the externally applied field $B_{\mathrm{ext}}$ from the effective field acting on the junction,
\begin{equation}
B_{\mathrm{eff}} = C\,B_{\mathrm{ext}} + B_{\mathrm{int}}.
\end{equation}
Here $C$ accounts for flux focusing and screening by the superconducting leads (see Sec.~S2 of Supplemental Material~\cite{SM}), while $B_{\mathrm{int}}$ denotes a configuration-dependent local-field offset associated with the metastable subsystem. In a magnetic interpretation, this offset can be parameterized as $B_{\mathrm{int}}=\kappa M$, where $M$ is an effective local magnetization and $\kappa$ encodes the coupling to the weak link. In this picture, the junction follows a smooth envelope as a function of $B_{\mathrm{eff}}$, while discrete metastable reconfigurations change $B_{\mathrm{int}}$, producing abrupt changes in both the shape and position of the pattern. This leads to an abrupt reorganization (shift and distortion) of the switching-current pattern when plotted versus $B_{\mathrm{ext}}$.

To assess the magnitude of the intrinsic local-field contribution implied by this picture, we use the expected value of the field of the first node of the Fraunhofer pattern, as extracted from the central lobe, $B_0\simeq 9.8$~mT (see Sec.~S3 of Supplemental Material~\cite{SM}). A change of the local offset by $\Delta B_{\mathrm{int}}$ corresponds to a flux shift $\Delta\Phi/\Phi_0 \approx \Delta B_{\mathrm{int}}/B_0$. Thus, even modest variations of the intrinsic configuration-dependent offset field $B_{\mathrm{int}}$ can induce sizable changes in $I_{\mathrm{sw}}$ when the envelope has a sufficiently large slope. This is consistent with the observation that the main discontinuity occurs at $|B|\simeq3$~mT, i.e., on the flank of the central lobe. 
In this regime, abrupt switching of magnetically active metastable configurations corresponds to an abrupt phase shift of the Fraunhofer pattern.

While the microscopic origin of this peculiar low-field response cannot be unambiguously identified, it may arise from collective reconfigurations of surface moments or correlated magnetic impurities in the hybrid NW environment, as discussed in related systems.~\cite{Strambini2020,Yang2021} 
Such reconfigurations can modify the local effective field and, consequently, the Andreev spectrum and the Josephson current of the weak link.~\cite{Bergeret2015} A minimal metastable model that produces avalanches with discrete jumps in an effective magnetization-like variable under quasi-adiabatic driving is provided in Sec.~S4 of Supplemental Material~\cite{SM}.

Finally, we assess several alternative mechanisms for the observed low-field switching, including a Zeeman-driven $0$--$\pi$ transition of the short weak link, local heating, charge-trap-induced transparency fluctuations, and rearrangements of the current distribution (see Sec.~S5 of Supplemental Material~\cite{SM}). Standard estimates for NW parameters place the Zeeman-driven transition field orders of magnitude above the observed $\sim$3~mT threshold, while the other mechanisms are disfavored as sole explanations of the sharp, reproducible, sweep-direction-dependent discontinuities.

In conclusion, we show that an $n$-doped InAs/Al NW JJ provides direct transport signatures of avalanche-like metastable reconfigurations, revealed as discrete switching-current jumps at perpendicular fields as low as $B_{\rm ext}\simeq 3$~mT. The sharp discontinuities, together with their reproducible sweep-direction dependence, are consistent with a Barkhausen-like phenomenology of avalanche switching in a magnetically active metastable degree of freedom coupled to the weak link. The nearly temperature-independent jump field, in contrast with the superconducting critical field, further disfavors simple superconducting mechanisms such as flux trapping as the sole origin of the observed discontinuities. Within an effective-field picture, each reconfiguration generates a discrete local-field offset, producing abrupt reorganizations of the switching-current response. These results establish hybrid NW JJs as sensitive probes of magnetically active avalanche-like metastability at low magnetic fields, and open new directions for exploiting metastable magnetic degrees of freedom in superconducting weak links in nanoscale devices.

\section{Acknowledgments}
We gratefully acknowledge S. Bergeret for fruitful discussions.
R. C. and C. G. acknowledge the PNRR MUR Project No. PE0000023-NQSTI (TOPQIN and SPUNTO). 
R. C. acknowledges Horizon Europe EIC Pathfinder under the Grant IQARO No. 101115190.
V. Z., L. S., E.S. and F.G. acknowledge the PNRR MUR Project PE0000023-NQSTI.
E.S acknowledges the partial support from the Italian National Research Council (CNR) under grant DFM.AD002.206 (HELICS)
A. B. acknowledges the project "Thermoelectric effects in solid-state quantum devices based on multiterminal Josephson junctions” of the bilateral agreement CNR/CONICET (Italy/Argentina) 2026-2027 and the CNR Project QTHERMONANO.

\section{Data Availability Statement}
The data are available from the authors upon reasonable request.

\section{Conflicts of Interest}
The authors declare no competing financial interest.

%\bibliographystyle{apsrev4-2}
%\bibliography{refs}


\clearpage
\onecolumngrid
\begin{center}
{\Large\bfseries Supplemental Material\par}
\medskip
{\large Switching-Current Signatures of Magnetically Active Avalanche-like Metastability in InAs/Al Nanowire Josephson Junctions\par}
\end{center}
\setcounter{figure}{0}
\setcounter{equation}{0}
\setcounter{table}{0}
\renewcommand{\thefigure}{S\arabic{figure}}
\renewcommand{\theequation}{S\arabic{equation}}
\renewcommand{\thetable}{S\arabic{table}}
\renewcommand{\theHfigure}{S\arabic{figure}}
\renewcommand{\theHequation}{S\arabic{equation}}
\renewcommand{\theHtable}{S\arabic{table}}
\section*{S1: Additional switching current patterns of different samples}

\begin{figure}[h!]
 \centering
 \includegraphics[width=\columnwidth]{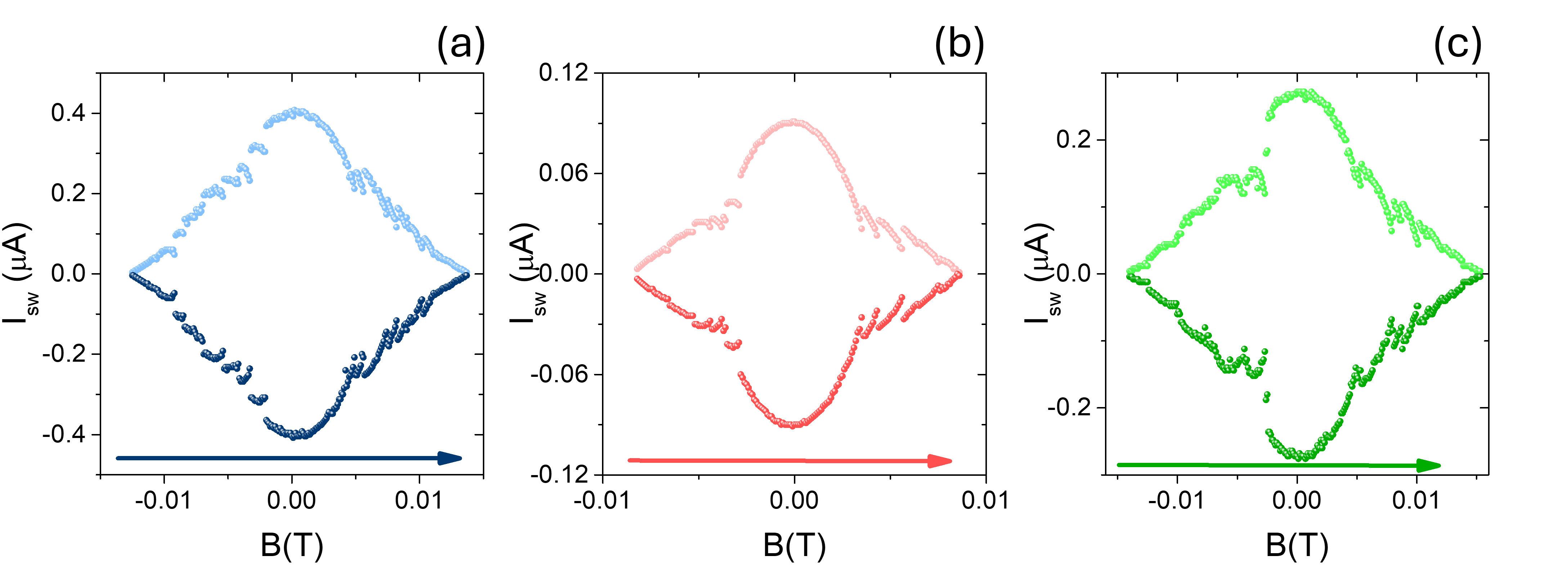}
 \caption{\textbf{Additional switching-current patterns across devices.} $I_{\rm sw}(B)$ for three devices: (a) D3-d5 at $30$ mK, (b) C2-d1 at $700$ mK, and (c) C2-d5 at $30$ mK. Data acquired with the magnetic field swept from negative to positive (see the arrow).}
 \label{fig:SI1}
\end{figure}

Figure~\ref{fig:SI1} (a–c) shows the switching current $I_{\rm sw} (B)$ as a function of out-of-plane magnetic field $B$ for three different InAs-nanowire (NW) Josephson junctions (JJs) fabricated and measured under comparable conditions: (a) Device D3-d5 measured at $T = 30\,\text{mK}$, (b) Device C2-d1 measured at $T = 700\,\text{mK}$, and (c) Device C2-d5 measured at $T = 30\,\text{mK}$. The colored arrows indicate the direction of the magnetic-field sweep.
Across all devices, $I_{\rm sw}(B)$ exhibits a lobe-shaped modulation (a central maximum with side lobes), consistent with a Fraunhofer-like interference pattern expected for flux-controlled Josephson transport. However, the devices share the same qualitative phenomenology: a central lobe followed, after the first jump at fields of a few mT, by a sequence of discontinuities that progressively reduce the switching current toward the field-driven suppression of the supercurrent. 
In addition, we observe abrupt jumps in $I_{\rm sw}$ at specific field values. As discussed in the main text and in other Supplementary sections, such jumps are naturally compatible with sudden changes of an effective intrinsic field acting on the junction, for instance due to discrete reconfigurations of a magnetically active metastable subsystem. This can manifest as small, abrupt shifts in the effective field and, more generally, as a reorganization (shift and distortion) of the interference pattern relative to the externally applied field. Finally, all the samples exhibit an asymmetry between the positive and negative branches with respect to the direction of the magnetic field sweep; the jumps are essentially temperature independent and remain in the same position across different sweeps.

\begin{figure}[h]
 \centering
 \includegraphics[width=0.5\columnwidth]{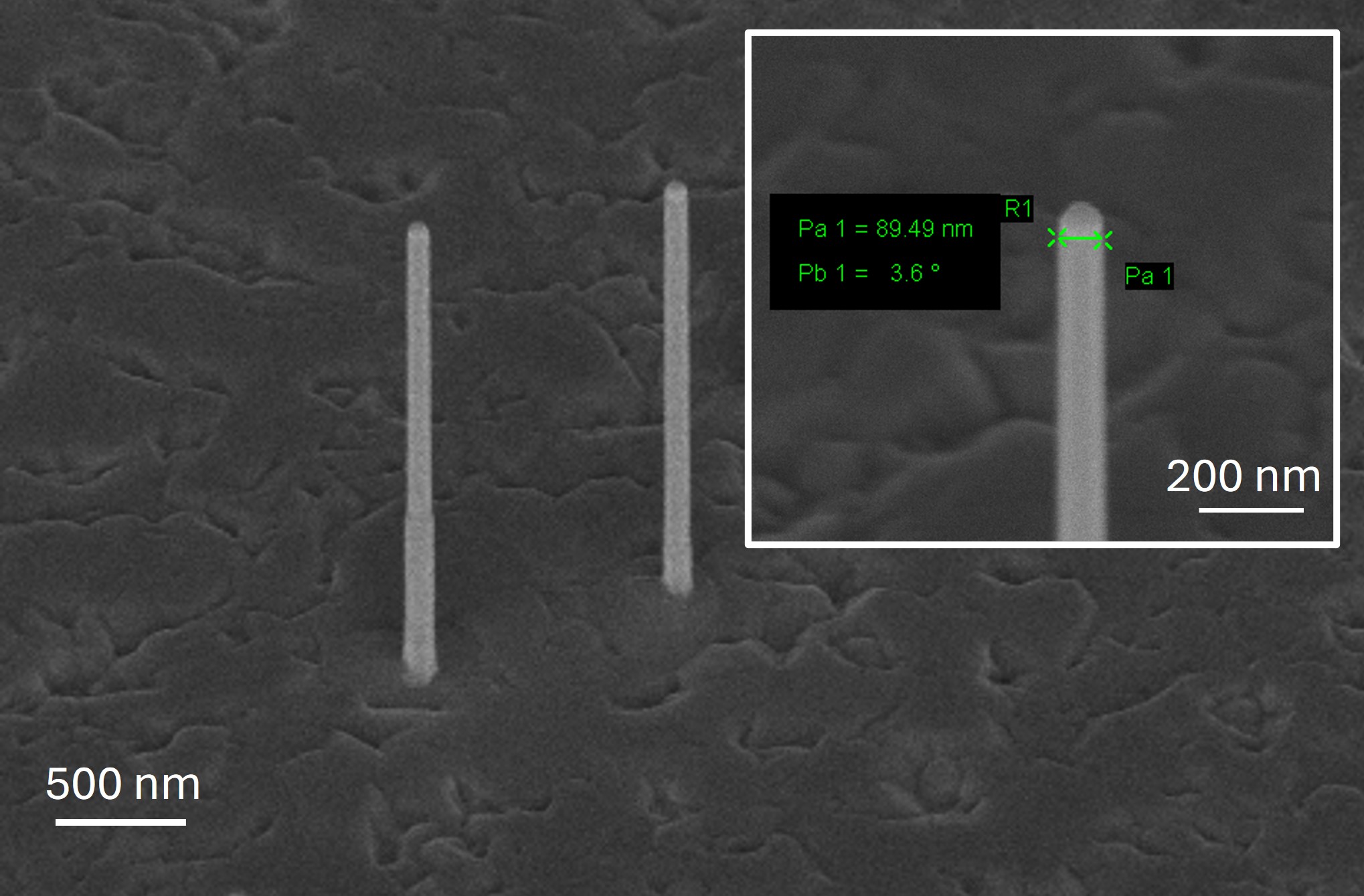}
 \caption{\textbf{45 degrees SEM micrograph of the InAs NW growth batch used in this work.}}
 \label{fig:SI2}
\end{figure}

Importantly, the coexistence of Fraunhofer-like lobes and switching-current jumps is reproducible across devices and temperatures, indicating that the observed phenomenology is not restricted to a single sample. While quantitative details vary from device to device, including the overall scale of $I_{\rm sw}(B)$ and the characteristic lobe spacing, as expected from differences in effective junction parameters, the qualitative behavior remains unchanged. In particular, the higher-temperature dataset in panel (b) shows a reduced $I_{\rm sw} (B)$ compared to the 30 mK measurements, consistent with the expected suppression of the supercurrent with increasing $T$. Finally, we note that the InAs NWs used to fabricate the different devices originate from the same growth batch (NW5498, Figure~\ref{fig:SI2}). This ensures comparable material quality and growth parameters, supporting the observed device-to-device reproducibility. At the same time, the observation of similar behavior within this batch should not be interpreted as evidence that such low-field metastable switching is universal across all InAs/Al NW JJs. Rather, it indicates that the effect is reproducible within the material and fabrication conditions explored in this work. Indeed, all devices successfully measured with the present technique show qualitatively similar low-field switching features, suggesting that the response is robust within this specific batch, while its broader generality across the InAs/Al platform remains to be established. Figure~\ref{fig:SI2} shows a representative scanning electron micrograph (SEM) overview of the as-grown NW ensemble: several NWs are visible within the same field of view with a largely similar morphology and dimensions. The image highlights the typical lengths of NWs in the $1 - 2\,\mu m$ range, while the inset highlights a representative NW diameter of 89~nm. The presence of multiple NWs with comparable diameter and length scale within the same SEM frame provides a direct visual reference for the growth batch used throughout this work.

\section*{S2: Flux-focusing estimate from electrode geometry}
\label{sec:SI_flux_focusing_geom}

In the presence of superconducting electrodes, demagnetization and screening currents concentrate magnetic field lines into the weak-link region (flux focusing). We parameterize this geometric enhancement by writing the local field in the junction as
\begin{equation}
B_{\rm eff}^{(0)} = C B_{\rm ext},
\label{eq:SI_Beff0}
\end{equation}
where $B_{\rm ext}$ is the externally applied out-of-plane magnetic field, $B_{\rm eff}^{(0)}$ is the effective field due to flux focusing/screening alone, i.e. without any additional local offset field discussed in Sec.~S4, and $C$ represents the focusing factor. The estimate below is intended to provide the relevant geometric field scale rather than a fitted parameter.

\begin{figure}[tbp]
 \centering
 \includegraphics[width=0.7\columnwidth]{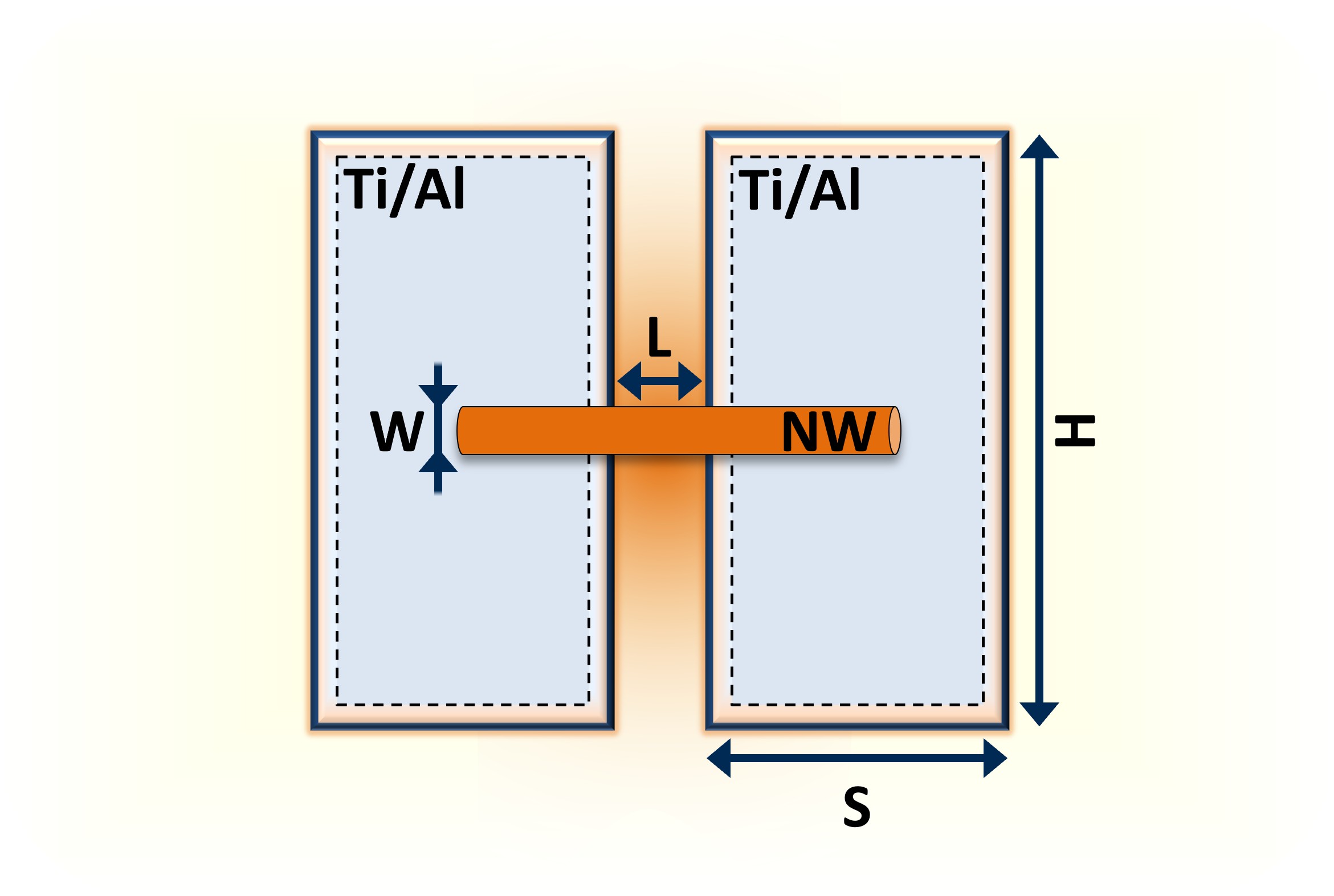}
 \caption{\textbf{Device geometry used for the flux-focusing estimate.}
Schematic top view of the SNS JJ. The superconducting Ti/Al leads (light blue) are characterized by lateral dimensions $S$ and $H$, while the NW weak link (orange, labeled NW) is characterized by width $W$ and length $L$. In the penetration-corrected estimate, the effective superconducting dimensions entering the focusing model are reduced to $S' = S - 2\lambda_{\rm thin}$ and $H' = H - 2\lambda_{\rm thin}$, while the effective weak-link length becomes $L' = L + 2\lambda_{\rm thin}$. The shaded orange contour indicates the region over which screening currents are assumed to collect flux.}
 \label{fig:SI3}
\end{figure}

We denote the weak-link dimensions by $W$ and $L$, and the superconducting lead dimensions by $S$ and $H$ (see Fig.~\ref{fig:SI3}). To include finite magnetic penetration into the superconducting leads, we introduce the effective dimensions
\begin{equation}
S' \approx S - 2\lambda_{\rm thin},
\qquad
H' \approx H - 2\lambda_{\rm thin},
\qquad
L' \approx L + 2\lambda_{\rm thin},
\label{eq:SI_effective_dims}
\end{equation}
where $\lambda_{\rm thin}$ is the effective magnetic penetration depth in the thin superconducting electrodes. The reduction of $S$ and $H$ accounts for the loss of fully screening superconducting area near the lead edges, while the increase of $L$ reflects field penetration into the leads at the weak-link interfaces.

Following the same simple geometric estimate for rectangular electrodes, the focusing factor can be approximated as~\cite{Amet2016,Villani2025,Clem2010}
\begin{equation}
C = 1 + \frac{2A_{\rm sc}}{L'H'},
\label{eq:SI_Cstar_def}
\end{equation}
where $A_{\rm sc}$ is the effective screening area of a single superconducting lead contributing to flux collection, which, for the rectangular geometry sketched in Fig.~\ref{fig:SI3}, is given by~\cite{Amet2016,Villani2025,Clem2010}
\begin{equation}
A_{\rm sc}
=
\left(\frac{S'}{2}\right)^2
+
\left(\frac{S'}{2}\right)\left(H'-S'\right)
=
\frac{S'}{2}\left(H'-\frac{S'}{2}\right),
\label{eq:SI_Asc}
\end{equation}
which yields
\begin{equation}
C
=
1+\frac{S'H'-S'^2/2}{L'H'}.
\label{eq:SI_Cstar_general}
\end{equation}
%
%For the present layout, where $H \simeq 2S$, this expression can be approximately written as
%
%\begin{equation}
%C \sim 1+\frac{3}{4}\frac{S'}{L'}.
%\label{eq:SI_Cstar_simplified}
%\end{equation}

Geometric parameters were estimated from SEM images and lithographic design. We take superconducting lead dimensions $S=(1000\pm5)$~nm and $H=(2000\pm5)$~nm, while the NW weak link has width $W=(80\pm5)$~nm and interelectrode spacing $L=(75\pm5)$~nm. 
For thin superconducting films, $\lambda_{\rm thin}$ can be expressed in terms of the bulk penetration depth $\lambda$ via the standard slab relation~\cite{LopezNunez2025IOP}
\begin{equation}
\lambda_{\rm thin} \approx \lambda\,\coth\!\left(\frac{d}{\lambda}\right).
\label{eq:SI_lambda_thin}
\end{equation}
For our Ti/Al leads, we take $d=100\pm5$~nm and a London penetration depth consistent with the range reported for nominally $d\simeq100$~nm Al films in Ref.~\cite{LopezNunez2025IOP}, i.e., $\lambda=(70\pm10)$~nm, which yields the value $\lambda_{\rm thin}\simeq 79\pm15$~nm. 
Substituting into Eq.~\eqref{eq:SI_Cstar_general}, and applying standard first-order error propagation to Eq.~\eqref{eq:SI_Cstar_general}, we obtain
\begin{equation}
C = 3.8 \pm 1.4,
\label{eq:SI_Cstar_value}
\end{equation}
where the quoted uncertainty corresponds to a $3\sigma$ confidence interval.
This penetration-corrected estimate is adopted below as the focusing factor for the discussion of the magnetic-field scales in the device. 

\section*{S3: Wide- and Narrow-Junction Limits}

As discussed by Cuevas and Bergeret~\cite{Cuevas2007PRL}, the magnetic-field dependence of the switching current in a diffusive junction can evolve from a Fraunhofer-like interference pattern to a monotonically damped response depending on the relation between the junction width and the magnetic length. 
In particular, the geometry-dependent crossover can be characterized by a magnetic length $\xi_H\sim\sqrt{\Phi_0/(2\pi B_{\rm eff})}$, where $B_{\rm eff}$ is the local field in the weak-link region.~\cite{Cuevas2007PRL,Montambaux2007ArXiv,Chiodi2012PRB}
Interference-induced nodes are expected when the transverse weak-link dimension exceeds this scale.~\cite{Cuevas2007PRL} For our junction, these criteria place the experiment close to an interference–dephasing crossover, so a Fraunhofer-like envelope with possible non-idealities (imperfect nodes and lobe distortions) is expected.

In order to compare these limiting cases, we introduce the dimensionless variable
\begin{equation}
x \equiv \frac{B+\delta B}{B_0},
\end{equation}
where $B_0$ is the characteristic magnetic-field scale (i.e., the quantity we are mainly interested in) and $\delta B$ may account for the small shift of the interference pattern with respect to zero applied field shown by our experimental data.
Using this notation, the wide-junction and narrow-junction switching-current responses are:
\paragraph{Wide junction, $W\gg\xi_H$~\cite{Cuevas2007PRL}:}
\begin{equation}\label{ISWwide}
I_{\mathrm{SW}}^{\mathrm{wide}}(B)
=
I_{\mathrm{SW}}^{\max}
\left|
\mathrm{sinc}\!\left( \pi x \right)
\right|
=
I_{\mathrm{SW}}^{\max}
\left[
1-\frac{1}{6}\left(\pi x\right)^2+\mathcal{O}(x^4)
\right],
\qquad x\to 0.
\end{equation}

\paragraph{Narrow junction, $W\ll\xi_H$~\cite{Montambaux2007ArXiv}:}
\begin{equation}\label{ISWnarrow}
I_{\mathrm{SW}}^{\mathrm{narrow}}(B)
=
I_{\mathrm{SW}}^{\max}
\frac{\left(\pi/\sqrt{3}\right)x}{
\sinh\!\left[\left(\pi/\sqrt{3}\right)x\right]}
=
I_{\mathrm{SW}}^{\max}
\left[
1-\frac{1}{6}\left(\frac{\pi}{\sqrt{3}}x\right)^2+\mathcal{O}(x^4)
\right],
\qquad x\to 0.
\end{equation}
Near zero field $x\ll 1$, the two expressions have the same qualitatively low-field structure, i.e., an inverted parabola, but they differ significantly away from the central maximum: the wide-junction response exhibits oscillatory lobes and exact zeros, whereas the narrow-junction response decays monotonically with field, see Fig.~3 of Ref.~\cite{Cuevas2007PRL}.
%As can be seen, the second-order expansion at small fields has the same structure in both cases, with the only difference in the prefactor, which determines the characteristic field scale from the fitting. 
In this sense, the narrow-junction response evolves more slowly with the field by a factor
$\sqrt{3} \approx 1.73$.
%At small fields, the two limiting responses have the same functional structure but different characteristic magnetic-field scales.

To compare the two functional forms on the same experimental scale, we fit the smooth low-field portion of the central lobe of the measured $I_{\rm sw}(B)$ using Eq.~\eqref{ISWwide}.
Fixing $I_{\mathrm{SW}}^{\max}=0.245~\mu\mathrm{A}$ to the maximum value in the dataset and leaving $B_0$ and $\delta B$ as free parameters, this fit gives
\begin{equation}
B_0^{\rm fit} = (9.8 \pm 0.8)\,\mathrm{mT},
\qquad\text{and}\qquad
\delta B^{\rm fit} = (0.5 \pm 0.2)\,\mathrm{mT},
\end{equation}
where the uncertainties quoted correspond to a $3\sigma$ confidence interval. 

\begin{figure}[t]
\centering
\includegraphics[width=0.75\linewidth]{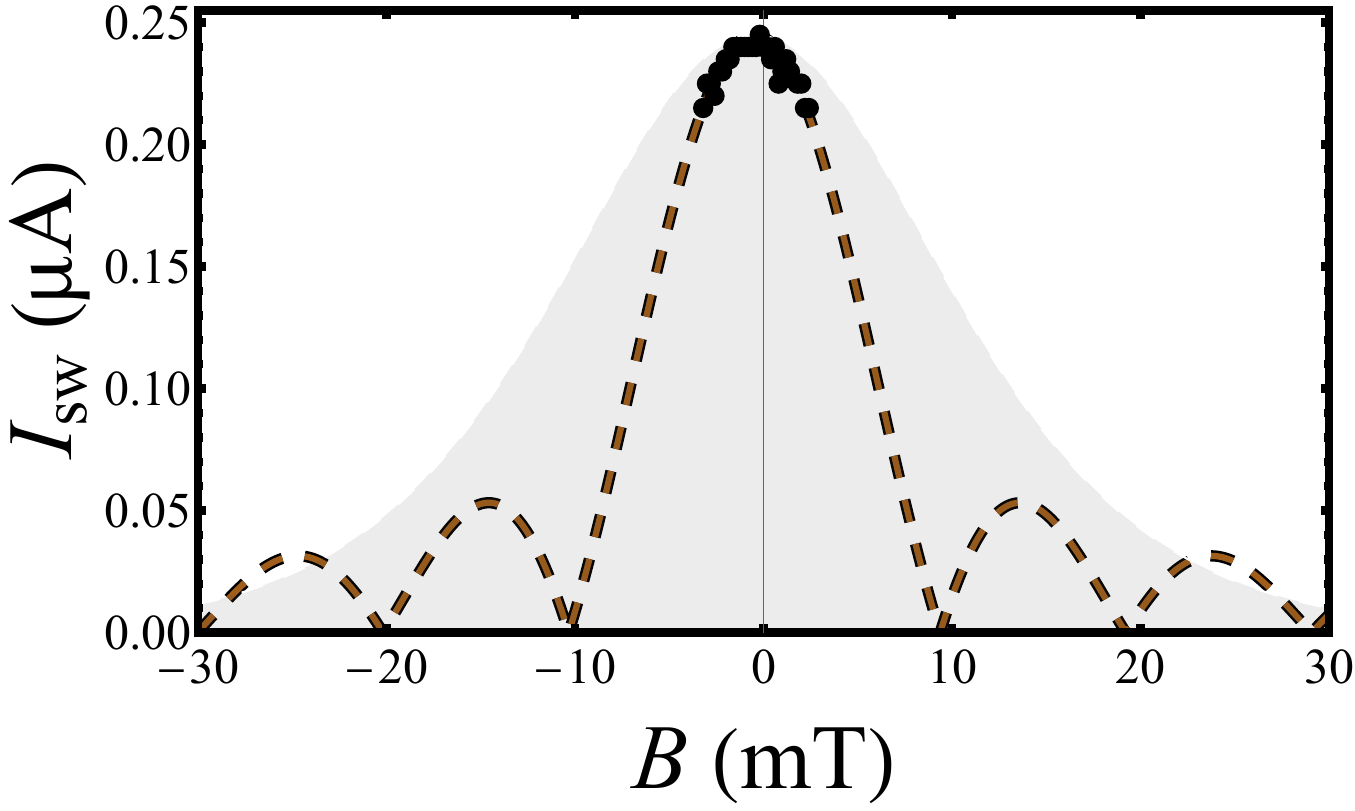}
\caption{\textbf{Comparison between the wide- and narrow-junction limiting forms of $I_{\mathrm{sw}}$ on the experimental magnetic-field scale.}
 Black dots: experimental data. Dashed line: fit to the data using Eq.~\eqref{ISWwide}. Light-gray shaded region: visual benchmark associated with the narrow-junction expression in Eq.~\eqref{ISWnarrow}, plotted using $(B_0^{\rm fit},\delta B^{\rm fit})$. The shaded area indicates the field range over which the narrow-junction limit is expected to suppress the switching current monotonically. This comparison highlights the difference between the oscillatory Fraunhofer-like response and the smoother narrow-junction decay.}
\label{fig:SI4}
\end{figure}

Figure~\ref{fig:SI4} shows this comparison. The black dots are the experimental data, and the colored dashed line is the fit obtained with Eq.~\eqref{ISWwide}. The light-gray shaded region is introduced as a visual reference for the narrow-junction limit, using Eq.~\eqref{ISWnarrow} with the same parameters extracted from the wide-junction fit. It is therefore meant as a benchmark of the expected monotonic field dependence in the narrow-junction case.

Thus, in realistic SNS junctions, the magnetic-field response is governed by the crossover between diffraction-like interference and orbital dephasing/depairing.~\cite{Cuevas2007PRL,Montambaux2007ArXiv,Chiodi2012PRB}
A practical criterion for the appearance of interference-induced lobes and nodes is that the transverse weak-link dimension exceeds the magnetic length, $\xi_H\sim\sqrt{\Phi_0/(2\pi B_{\rm eff})}$.~\cite{Cuevas2007PRL}
Using the penetration-corrected effective field $B_{\rm eff}\simeq C B_{\rm ext}$ and the effective junction length $L' = L + 2\lambda_{\rm thin}$, this criterion defines a crossover field scale
\begin{equation}
B_{\rm ext}^{\rm cross}\sim \frac{\Phi_0}{2\pi\,C W L'}.
\end{equation}
With $W\simeq 80$~nm, $L' = 233~\mathrm{nm}$, and $C = 3.8 $, we obtain
\begin{equation}
B_{\rm ext}^{\rm cross}= (4.6\pm1.7)~\mathrm{mT}
\qquad (3\sigma).
\end{equation}

This estimate is especially relevant because the magnetic-field range explored experimentally extends well into the regime $|B|\gtrsim B_{\rm ext}^{\rm cross}$. Therefore, our devices are expected to develop a pronounced lobe structure over a substantial field interval, far from a purely monotonic narrow-junction regime.

At the same time, the junction geometry is not deep in the wide-junction limit. Cuevas \emph{et al.} remark that the standard Fraunhofer pattern is approached when $W\gtrsim L$.~\cite{Cuevas2007PRL} In our devices a Fraunhofer-like pattern is just roughly expected being $W\simeq 80$~nm and $L\simeq 70$--$75$~nm. Indeed, this places the junction in an intermediate-width regime, where interference lobes are expected to persist, but the full pattern need not coincide with the standard Fraunhofer form for a uniform rectangular junction. In the same way, Chiodi \emph{et al.} report a clear lobe structure in a diffusive SNS junction with $W/L\simeq 1.4$ (see Fig.~3 of Ref.~\cite{Chiodi2012PRB}), showing experimentally that lobe patterns persist even outside the extreme wide-junction limit.

Overall, these considerations suggest that our junction operates in a regime where the magnetic-field response should remain substantially lobed, yet generally non-Fraunhofer: the side lobes may vary in height, width, and position, and the first minimum is not expected to coincide exactly with the ideal Fraunhofer node.~\cite{Cuevas2007PRL,Chiodi2012PRB} This is qualitatively consistent with the experimental $I_{sw}(B)$ patterns, which retain a non-monotonic lobe structure while showing clear deviations from a standard Fraunhofer response.
For this reason, fitting only the central lobe with a $|\mathrm{sinc}|$ form should be regarded as a local parametrization around $\Phi=0$, providing an effective low-field scale for the initial suppression of the switching current rather than a quantitatively exact prediction of the first-node position.

From the effective flux through the weak link, by imposing $\Phi\simeq C B_{\rm ext}\,W L'=\Phi_0$, we obtain a geometric \emph{estimated first-node field}
\begin{equation}
B_0^{\rm est} \approx \frac{\Phi_0}{C W L'} \approx (29.2\pm5.6)~\mathrm{mT}\qquad (3\sigma).
\end{equation}
As expected, this estimate is larger than the fitted value, $B_0^{\rm fit}$, which is based on an idealized Fraunhofer-like geometric argument for the first node, while our junction lies in an intermediate regime, close to the crossover between wide- and narrow-junction behavior. In this situation, the inner part of the central lobe can still be locally described in first approximation by Eq.~\eqref{ISWwide}, while the full magnetic-field response is not expected to follow the standard Fraunhofer pattern quantitatively. For this reason, $B_0^{\rm fit}$ should be regarded as an effective low-field scale for the initial suppression of the switching current, rather than as a direct geometric prediction of the first node position.~\cite{Cuevas2007PRL,Chiodi2012PRB}

\section*{S4: Minimal random-field Ising model for avalanche-like metastable reconfigurations}

Here we provide a compact lattice-model picture for the step-like changes of the junction response observed under quasi-adiabatic magnetic-field sweeps. The model describes discrete reconfigurations of a magnetically active metastable subsystem and is used here as a minimal proxy for field-driven metastable switching events that can underlie the observed $I_{\mathrm{sw}}(B)$ jumps. Our goal is not a fully quantitative micromagnetic description, which would require a clearly identified microscopic origin of the magnetically active subsystem, but rather a minimal framework that shows how a finite number of effective metastable regions can generate discrete switching events (``\emph{avalanches}'') and a metastable response under slow magnetic driving.
We stress that the RFIM variables used here should not be interpreted as directly imaged magnetic domains with an independently measured size. Unlike EuS-based hybrids, where engineered ferromagnetic layers allow magnetic textures to be correlated with superconducting transport and with the superconducting coherence-length scale,~\cite{Nandi2026,Maiani2025} our InAs/Al devices, in principle, do not contain intentionally introduced magnetic layer and no magnetic imaging was directly available. The \emph{random-field Ising model} (RFIM) should therefore be regarded as a minimal coarse-grained description of avalanche-like metastability, with the effective spins representing local metastable configurations of a magnetically active subsystem.
Minimal models of hysteresis and crackling noise based on metastable RFIM dynamics have a long history in the context of Barkhausen-like avalanches and related disorder-driven switching phenomena~\cite{Barkhausen1919,Sethna2001,Sethna1993,Dahmen1993,Dahmen1996,Perez2003,Bertotti1998,Zapperi2001,Durin2006}. In this class of models one considers Ising spins $s_i\in\{-1,+1\}$ on a rectangular lattice $N=L_xL_y$ sites and Hamiltonian
\begin{equation}
\mathcal{H}_0
=
-J\sum_{\langle i,j\rangle} s_i s_j
\;-\;
\sum_i (H_{\mathrm{ext}} + h_i)\, s_i,
\end{equation}
where $J>0$ is the ferromagnetic nearest-neighbor coupling, $H_{\mathrm{ext}}$ is the externally controlled driving field, and $h_i$ is a \emph{quenched} random field drawn once and kept fixed throughout the entire sweep. The normalized magnetization is
\begin{equation}
M(H_{\mathrm{ext}})
=
\frac{1}{N}\sum_{i=1}^{N} s_i.
\end{equation}
Here and in the following, we denote the driving field in the lattice model by $H_{\mathrm{ext}}$, to be identified phenomenologically with the experimental out-of-plane field $B_{\mathrm{ext}}$ up to a proportionality factor.

We assume zero-temperature ($T=0$) metastable dynamics: for a given value of $H_{\mathrm{ext}}$, the system is relaxed until all spins are locally stable. For a spin at site $i$, the local effective field is
\begin{equation}
f_i
=
J\sum_{j\in nn(i)} s_j + H_{\mathrm{ext}} + h_i,
\end{equation}
where $nn(i)$ denotes nearest neighbors. The dynamics is deterministic: a spin is locally stable if $s_i=\mathrm{sign}(f_i)$, with the convention that no flip occurs at exact degeneracy $f_i=0$, and unstable otherwise, in which case it flips toward $\mathrm{sign}(f_i)$. A quasi-adiabatic sweep is implemented by incrementing $H_{\mathrm{ext}}$ in small steps and fully relaxing the system at each step. A single flip may trigger a cascade of further flips because the local fields of neighboring sites are modified after every update. 
The avalanche size is the total number of flips occurring during one relaxation step and determines the corresponding discrete jump in $M(H_{\mathrm{ext}})$.

For the quenched disorder, we use a phenomenological description in which the magnitude of the local anisotropy field is drawn from a normal distribution,
\begin{equation}
p(|h|)=\mathcal{N}(\mu_h,\sigma_h^2),
\end{equation}
where $\mu_h$ sets the characteristic switching scale and $\sigma_h$ its dispersion, while the sign of $h$ for each effective spin is chosen randomly. 
Compared with the standard Gaussian RFIM, this choice introduces a characteristic anisotropy scale, set by $\mu_h$, with a spread controlled by $\sigma_h$. Such choices for the random-field distribution have been discussed in the RFIM literature~\cite{Fytas2008a,Fytas2008b,Hernandez2008,Hadjiagapiou2010,Akinci2011,Fytas2013,Fytas2016}. 
In this coarse-grained description, the typical anisotropy scale sets the threshold field required to trigger the first avalanche-like jump. A quantitative microscopic prediction of this scale is not attempted here, because the present transport data do not uniquely identify the microscopic origin, volume, or correlation length of the metastable magnetic subsystem.

\begin{figure}[t]
  \centering
  \includegraphics[width=0.9\linewidth]{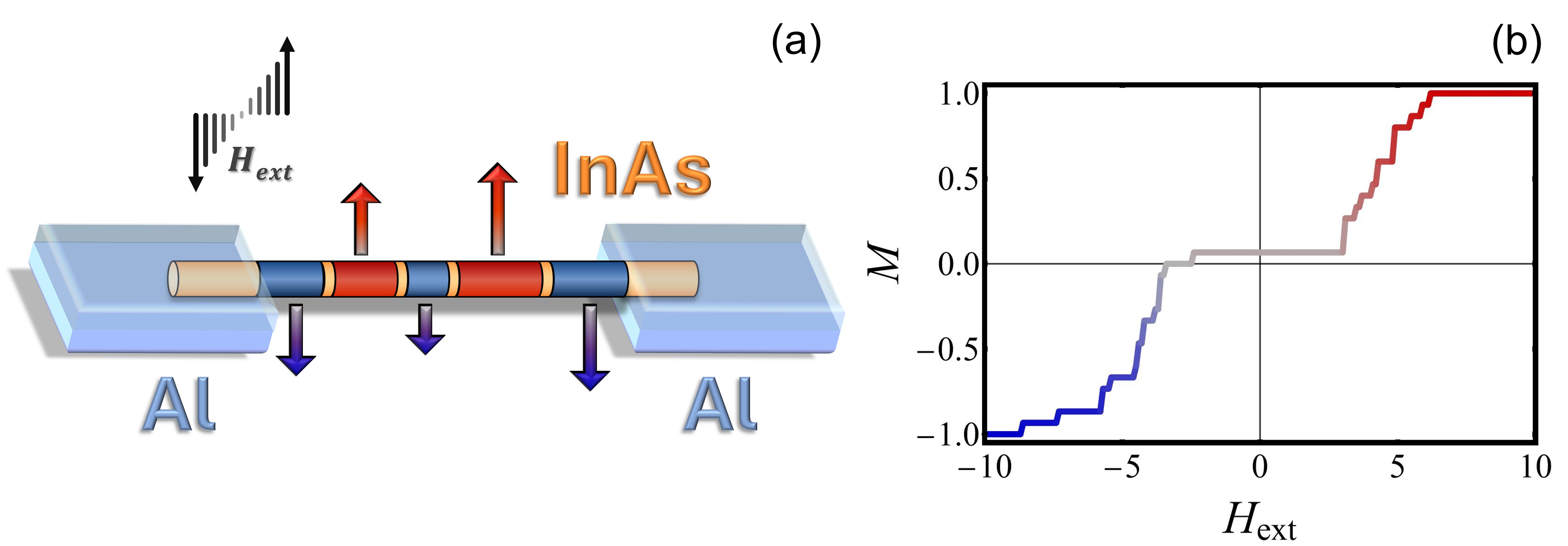}
  \caption{\textbf{Magnetization arrangement and representative RFIM output.}
  (a) Cartoon of the InAs NW JJ between two superconducting Al leads driven by a quasi-adiabatic external magnetic-field sweep. The NW is represented as a sequence of effective magnetically active metastable regions (blue/red), corresponding to two local configurations of the coarse-grained RFIM variables.
  (b) Example magnetization curve $M(B)$ from a quasi-adiabatic RFIM strip, $L_x\times L_y=15\times 2$ and parameters: $J=0.5$, $\mu_h=5$, and $\sigma_h=1.5$.}
  \label{fig:SI5}
\end{figure}

To mimic a wire-like magnetic element, we use a long-and-narrow strip lattice with $L_x\gg L_y$. This choice is also motivated by the fact that elongated geometries make discrete switching events more visible: in nearly square lattices, the magnetization response typically fragments into many small steps, while in strip-like geometries one more readily obtains sharper avalanche-like jumps. In the simulations shown in Fig.~\ref{fig:SI5}, we adopt \emph{cylindrical} boundary conditions, periodic along the transverse direction $y$ and open along the longitudinal direction $x$, and initialize the system in a negatively saturated state, consistent with an experimental protocol that starts from large negative fields and sweeps toward positive fields.
We also verified that, for the strip-like geometries considered here, the same qualitative phenomenology is obtained with fully open boundaries, indicating that the emergence of the discussed physics is not critically tied to the specific boundary-condition choice.
The lattice size and RFIM parameters are chosen to illustrate avalanche-like metastable switching in a finite strip and should not be used to infer a microscopic domain size.

Figure~\ref{fig:SI5} summarizes the minimal picture used here. Panel (a) shows a schematic view of the InAs NW JJ, where the magnetically active region is represented as a sequence of coarse-grained metastable configurations. This cartoon is not meant as a microscopic reconstruction of the actual domain texture, but rather as a minimal representation of a finite magnetically active metastable system that can undergo discrete field-driven reconfigurations.

Panel (b) shows a representative magnetization curve $M(H_{\mathrm{ext}})$ obtained from the corresponding RFIM strip under a quasi-adiabatic field sweep. The response is step-like, reflecting discrete avalanches that connect successive metastable states as the driving field is varied. 
In this sense, the model provides a minimal picture of how a finite magnetically active metastable subsystem can generate abrupt switching events and metastable threshold-like behavior under slow magnetic driving.
At the same time, the central part of the curve may display a low-magnetization plateau that is not necessarily exactly centered at $M=0$. Within the effective-field picture discussed below, such a small bias in $M$ naturally translates into a rigid shift of the corresponding interference pattern and therefore can provide a phenomenological interpretation of the offset parameter $\delta B$ introduced in Sec.~S3 for the central-lobe fit. This is also consistent with the small low-field shifts visible in the additional critical-current patterns of Fig.~S1.

The visibility of such jumps is enhanced in narrow strip-like geometries for two related reasons. First, finite-size normalization makes each avalanche more visible: an avalanche of size $s$ produces a magnetization jump $\Delta M \simeq 2s/(L_xL_y)$. Second, elongated geometries with $L_x\gg L_y$ tend to favor more correlated propagation along the long direction, whereas in more isotropic or nearly square shape the response typically fragments into many smaller steps that are less sharply resolved. This is why the strip geometry is not only a natural proxy for a wire-like magnetic element, but also a convenient regime in which discrete avalanche-like jumps in $M(H_{\mathrm{ext}})$ are more clearly observable.

Importantly, for a fixed realization of the quenched disorder and a fixed sweep protocol, the model produces a reproducible sequence of threshold-like jumps in $M$ together with a low-magnetization central plateau. This is a direct consequence of the deterministic $T=0$ metastable dynamics.
In particular, the central plateau may occur at small values of $M$, including $M=0$, consistent with the idea that different metastable coarse-grained configurations can produce similar qualitative behavior, together with different effective offsets in the junction response. What matters in this regime is not that $M$ vanishes exactly, but that it remains nearly field-independent over a finite interval, so that its contribution acts mainly as a rigid offset of the interference pattern until the first jump at $B_j^\pm$.

In a mesoscopic device, the relevant field acting on the weak link is generally not the uniform applied field alone, but an effective local field that also includes a contribution from configuration-dependent local magnetic offsets. A minimal phenomenological relation can be written as
\begin{equation}
B_{\mathrm{eff}}=C\,B_{\mathrm{ext}} + \kappa\, M(B_{\mathrm{ext}}),
\end{equation}
where $C$ accounts for flux focusing or screening by the superconducting leads (see Section~S2), while $\kappa M$ represents the configuration-dependent local field contribution generated by the nearby magnetically active metastable subsystem.
In this form, discrete jumps in $M$ translate directly into discrete changes of the effective field experienced by the junction:
\begin{equation}
\Delta B_{\mathrm{eff}} = \kappa\,\Delta M.
\end{equation}
A Barkhausen-like, avalanche-type reconfiguration in this coarse-grained model therefore shifts the effective flux threading the junction and can appear experimentally as an abrupt horizontal displacement of the Fraunhofer-like pattern, or more generally as an abrupt reorganization of the underlying Fraunhofer-like pattern, which appears experimentally as jumps in $I_{\mathrm{sw}}(B)$.
Within the same effective-field picture, a small quasi-constant bias of $M$ over the central field range simply produces a rigid offset of the central lobe of the pattern. 
In our data, small low-field offsets of the central lobe are consistently observed, both in the main dataset through the fitted parameter $\delta B$ and in the additional device traces shown in Fig.~S1.

\section*{S5: Alternative mechanisms for the low-field switching}

In this section we discuss alternative mechanisms that could, in principle, produce changes in the measured switching current of an inhomogeneous semiconductor weak link. We focus on four possibilities: a Zeeman-driven $0$--$\pi$ transition, local heating, charge-trap-induced transparency changes, and rearrangements of the supercurrent distribution. The goal is not to exclude every possible microscopic source of metastability, but to assess whether these mechanisms naturally account for the combined experimental observations: reproducible jumps at well-defined low magnetic fields, pronounced dependence on the magnetic-field sweep direction, weak temperature dependence of the jump fields compared with the superconducting critical field, and coexistence of the jumps with a Fraunhofer-like modulation of $I_{\mathrm{sw}}(B)$.

\paragraph*{Zeeman-driven $0$--$\pi$ transition.}

An alternative mechanism to consider is a Zeeman-driven $0$--$\pi$ transition of the NW JJ.
Such Zeeman-induced nodes/minima of the Josephson response have been discussed in Dirac-semimetal-based junctions with large effective $g$ factors~\cite{Li2019Dirac}
and in planar InAs/Al junctions under strong in-plane fields.~\cite{Banerjee2023,Monroe2024}
In the short-junction limit, a standard ballistic estimate for the field scale of the first Zeeman-driven node/minimum
(often denoted $B_{\mathrm{FFLO}}$ or $B_{\mathrm{node}}$) is~\cite{Dartiailh2021PRL,Hart2017NatPhys}
\begin{equation}
B_{Z} \simeq \frac{\pi}{2}\frac{\hbar v_{F}}{L g \mu_{B}},
\label{eq:BZ_ballistic}
\end{equation}
where $v_{F}$ is the Fermi velocity, $L$ the junction length, $g$ the effective Land\'e factor, and $\mu_{B}$ the Bohr magneton. 
Using parameter values reported for nominally identical NW junctions in Ref.~\cite{Strambini2020}, that is $v_F\simeq 2\times 10^{6}$~m/s and $g=10$, if $L=70$~nm one obtains $B_Z \simeq 50~\mathrm{T}$, i.e., more than four orders of magnitude larger than the characteristic external field scale of our switching ($\sim 3$~mT).
Even if one takes the full NW length as an upper bound for the relevant traversal length ($L\sim 2~\mu$m),
Eq.~(\ref{eq:BZ_ballistic}) still yields $B_Z\sim 2~\mathrm{T}$, far above the millitesla regime. 
Similarly, even assuming $g$ factors larger by an additional order of magnitude would still leave the estimated Zeeman node field well above the observed low-field switching scale.
\paragraph*{Local heating.}

Self-heating is a well-established source of hysteresis in superconducting weak links and proximity JJs. In SNS junctions, the transition to the dissipative branch can increase the electron temperature in the normal region, thereby lowering the retrapping current and producing hysteretic current-voltage characteristics.~\cite{Courtois2008PRL,Li2011PRB} This mechanism is therefore relevant for understanding the difference between switching and retrapping currents in our $V$--$I$ curves. However, it is unlikely to be the sole origin of the low-field discontinuities reported here. A heating-induced instability is primarily controlled by dissipated power, bias-current history, and thermal relaxation, rather than by a reproducible threshold in the externally applied magnetic field. In our measurements, the jumps occur at characteristic field values during quasi-adiabatic field sweeps and are superimposed on an otherwise smooth low-field interference envelope. Moreover, the jump fields remain nearly unchanged while the overall superconducting scale and the critical field vary strongly with temperature. A purely thermal mechanism would be expected to track the temperature-dependent superconducting and thermal-relaxation scales more closely, and to produce broader or more stochastic switching features. We therefore regard local heating as disfavored as the sole origin of the sharp, history-dependent low-field jumps.

\paragraph*{Charge traps and junction-transparency fluctuations.}

Charge traps or two-level fluctuators near the NW, at the dielectric interface, or in the contact environment can modulate the electrostatic potential and hence the transparency of a semiconductor Josephson weak link. Similar fluctuators are known to generate critical-current noise in JJs, and microscopic models based on two-level systems or trapping centers have been extensively discussed in the context of superconducting tunnel junctions and qubits.~\cite{Faoro2007PRB,Constantin2007PRL,DeSousa2009PRB} In our devices, such mechanisms could contribute to background switching-current fluctuations or device-to-device variability. Nevertheless, a charge-trap scenario does not naturally explain the systematic magnetic-field phenomenology observed here. A trap-induced change of junction transparency would primarily rescale the magnitude of $I_{\mathrm{sw}}$ or produce telegraph-like temporal fluctuations, whereas the observed discontinuities occur at reproducible magnetic-field thresholds, depend on the direction of the field sweep, and appear as a reorganization of a field-dependent interference pattern. To account for these features, a charge trap would itself need to be coupled to the magnetic-field history, effectively reintroducing a field-driven metastable degree of freedom. We therefore consider charge-trap-induced transparency changes unlikely to provide the primary explanation for the observed low-field jumps.

\paragraph*{Rearrangements of the current distribution.}

A further possibility is that the supercurrent density in the weak link is spatially inhomogeneous and rearranges as the magnetic field is swept. Static inhomogeneity of the current distribution is well known to distort Josephson interference patterns: the measured $I_c(B)$ response encodes the spatial distribution of supercurrent density, and nonuniform current flow can lead to nonideal lobe amplitudes, shifted minima, and deviations from a textbook Fraunhofer pattern.~\cite{Dynes1971PRB,Fermin2023PRB} Such effects are certainly relevant for interpreting the nonideal Fraunhofer-like envelope in our devices. However, static inhomogeneity alone produces a distorted but continuous magnetic-field dependence. Abrupt, reproducible jumps require an additional bistable or multistable internal variable that changes configuration during the field sweep. If this variable were purely superconducting in origin, for instance associated with vortex entry, flux trapping, or local superconducting weak spots in the leads, its characteristic field scale would be expected to follow superconducting length and field scales, such as the penetration depth, coherence length, or critical field. Moreover, vortex entry, trapping, and detrapping in mesoscopic superconducting leads are typically sensitive to microscopic pinning conditions, residual trapped flux, and sweep history, and would therefore be expected to produce stronger scan-to-scan variability in the jump positions and amplitudes. This contrasts with the reproducible low-field discontinuities observed here, whose threshold fields remain nearly unchanged over repeated sweeps and only weakly depend on temperature. This is difficult to reconcile with the robust weak temperature dependence of $B_j^\pm$ over a temperature range where the superconducting critical field changes substantially. We therefore conclude that current-density inhomogeneity may contribute to the detailed shape of the interference pattern, but is unlikely by itself to generate the observed sharp, sweep-direction-dependent low-field discontinuities.

Taken together, these considerations disfavor a Zeeman-driven $0$--$\pi$ transition, local heating, charge trapping, and smooth current-density redistribution as sole explanations of the data. The observations are more naturally captured by a metastable degree of freedom that is both coupled to the weak link and driven by the magnetic-field sweep. In view of the independent evidence for localized magnetic moments and anomalous Josephson phase shifts in nominally identical InAs/Al NWs, a magnetically active metastable subsystem provides a natural microscopic interpretation.

%\bibliography{refs}

\end{document}